\documentclass[a4paper,11pt]{article}
\pdfoutput = 1
\usepackage{jheppub}
\usepackage{amsmath,amssymb,amscd,braket,amsfonts}
\usepackage{color}
\usepackage{graphicx}
\usepackage{slashed}
\usepackage{amsthm}
\usepackage{verbatim}
\usepackage{relsize}

\newcommand{\bea}{\begin{eqnarray}}
\newcommand{\eea}{\end{eqnarray}}
\newcommand{\be}{\begin{equation}}
\newcommand{\ee}{\end{equation}}
\newcommand{\ba}{\begin{align}}
\newcommand{\ea}{\end{align}}

\title{
Black hole excited states from broken translations in Euclidean time
}
\date{2023}

\author[a]{Christiana Pantelidou}
\affiliation[a]{School of Mathematics and Statistics, University College Dublin, Belfield, Dublin 4, Ireland}
\emailAdd{christiana.pantelidou@ucd.ie}

\author[b]{and Benjamin Withers}
\affiliation[b]{Mathematical Sciences and STAG Research Centre, University of Southampton, Highfield, Southampton SO17 1BJ, UK}
\emailAdd{b.s.withers@soton.ac.uk}

\abstract{
We prepare an excited finite temperature state in ${\cal N}=4$ SYM by means of a Euclidean path integral with a relevant deformation. The deformation explicitly breaks imaginary-time translations along the thermal circle whilst preserving its periodicity. We then study how the state relaxes to thermal equilibrium in real time.
Computations are performed using real-time AdS/CFT, by constructing novel mixed-signature black holes in numerical relativity corresponding to Schwinger-Keldysh boundary conditions. These correspond to deformed cigar geometries in the Euclidean, glued to a pair of dynamical spacetimes in the Lorentzian. 

The maximal extension of the Lorentzian black hole exhibits a `causal shadow', a bulk region which is spacelike separated from both boundaries. We show that causal shadows are generic in path-integral prepared states where imaginary-time translations along the thermal circle are broken. 
}

\begin{document}
\maketitle

\section{Introduction}

Holography provides a non-perturbative framework to study strongly coupled theories, such as ${\cal N} = 4$ SYM at a large $N$, through calculations in general relativity in asymptotically AdS spacetimes \cite{Maldacena:1997re}.
In this paper we use holography to study excited states of thermal systems in real time, and study their approach to thermal equilibrium.
The main motivation is to further develop and explore real-time holographic frameworks such as \cite{Skenderis:2008dh, Skenderis:2008dg, Glorioso:2018mmw}, and to better understand the exotic class of mixed-signature black hole spacetimes that compute real-time QFT observables.

Lorentzian evolution in AdS gravity is an initial-boundary-value problem, requiring both initial data and boundary data.
For excited states, the immediate question one must address is how to choose the initial data. 
There are two common approaches taken in numerical relativity.
One approach starts with initial data for vacuum or thermal equilibrium and performs a quench, where QFT sources vary in time and drive the system out of equilibrium.
The other approach taken is to make an ad-hoc choice of initial data. In the latter approach it is not clear precisely how this choice of initial data is encoded in the QFT, or indeed whether this is a valid QFT computation at all \cite{Belin:2020zjb}.

\begin{figure}[h!]
\centering
\includegraphics[width=0.5\columnwidth]{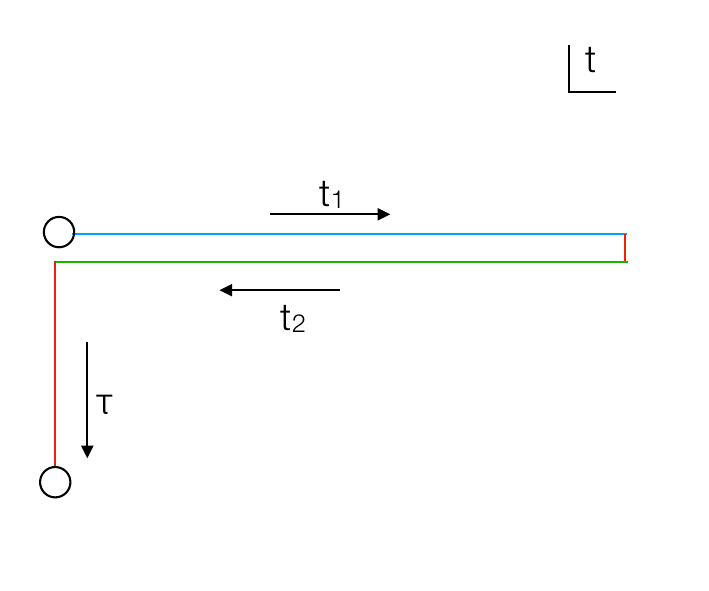}
\put(-150,113){$\mathcal{C}_1$}
\put(-105,87){$\mathcal{C}_2$}
\put(-194,80){$\mathcal{C}_E$}
\caption{The Schwinger-Keldysh contour ${\cal C}$ in the complex time plane.
The circles are identified to give a closed path. The associated Euclidean periodicity is the inverse temperature $\beta=1/T$. For each segment $t$ is parameterised as follows; on ${\cal C}_1$ we have $t=t_1$ with $t_1\in [0,t_f]$ for some sufficiently large $t_f>0$, on ${\cal C}_2$ we have $t=-t_2$ with $t_2\in [-t_f,0]$, and on ${\cal C}_E$ we have $t = -i \tau$ with $\tau \in [0, \beta]$.}
\label{fig:SK_contour}
\end{figure}

In this work we consider the QFT on a closed path in the complex time plane, the Schwinger-Keldysh contour ${\cal C}$, see figure \ref{fig:SK_contour}.
The holographic dual is no longer an initial-boundary-value problem, but a boundary-value problem with a mixed signature \cite{Skenderis:2008dh, Skenderis:2008dg}.
Consequently, initial data is no longer an input to the calculation. Instead, the task is to find a bulk saddle point geometry whose boundary is in the conformal class of the Schwinger-Keldysh contour itself.
In this context one can generate coherent excited states by turning on finite sources for single-trace operators in the Euclidean segment ${\cal C}_E$.

The present approach comes with two marked advantages.
First, by also considering perturbative sources on ${\cal C}_1$ and ${\cal C}_2$, one can compute a variety of real-time observables such as retarded, advanced and symmetric two-point functions, as well as analogous sets of higher-point functions. For example, retarded thermal three-point functions were recently computed using these techniques \cite{Loganayagam:2022zmq, Pantelidou:2022ftm}.
Second, the computation is defined in QFT terms, requiring no bulk-intrinsic input such as ad-hoc choices of initial data to define the state, or ingoing boundary conditions at horizons to compute correlators \cite{vanRees:2009rw}.

Related work preparing coherent excited states in vacuum includes \cite{Botta-Cantcheff:2015sav, Christodoulou:2016nej, Botta-Cantcheff:2017qir, Marolf:2017kvq, Belin:2020zjb}. Specifically,  \cite{Botta-Cantcheff:2015sav} proved that turning on non-trivial Euclidean sources gives rise to coherent bulk states, while \cite{Christodoulou:2016nej} and \cite{Botta-Cantcheff:2017qir} considered the effect of multi-trace operators and backreaction, and the inclusion of interactions, respectively.  Subsequently, \cite{ Marolf:2017kvq, Belin:2020zjb} investigated the extend to which arbitrary bulk coherent states can be represented by such Euclidean path-integrals in the CFT. Going beyond vacuum, \cite{Botta-Cantcheff:2018brv, Botta-Cantcheff:2019apr, Chen:2019ror} showed that sources on the closed SK contour correspond to thermal coherent states while thermalisation was discussed in \cite{Martinez:2021uqo}. \cite{Arias:2020qpg} studied the Modular Hamiltonian and Modular Flow for such bulk coherent states. Real time black hole formation in AdS$_3$ was studied in the in-in formalism in \cite{Anous:2016kss} by an excitation of the vacuum, prepared in the Euclidean by inserting a large number of primary operators on a circle at fixed radius.

In more detail, the QFT computation we are performing this work is as follows.
We consider ${\cal N}=4$ SYM with gauge group $SU(N)$ at large $N$.
This theory contains the following $\Delta = 3$ operator, the mass term for the gaugino $\lambda^4$,
\be
O_3 \sim \text{tr}(\lambda^4 \lambda^4) + h.c.
\ee
We consider the path integral on the Schwinger-Keldysh contour ${\cal C}$, and introduce sources $\lambda_{\cal C}$ for the operator $O_3$ along this contour,
\be
Z_{\text{CFT}}[\lambda_{\cal C}] = \int {\cal D}\chi \exp\left(-i S_{\text{CFT}} -i \int_{\cal C} d^4x\lambda_{\cal C}(x) O_3(x) \right), \label{pathint}
\ee
where $\chi$ collectively denote the CFT fields. 
The source function is comprised of three pieces; on the Euclidean segment ${\cal C}_E$ we write $\lambda_{\cal C} = \lambda(\tau)$ while on the Lorentzian segments ${\cal C}_{1,2}$ we write $\lambda_{\cal C} =J_{1,2}$ respectively.
To prepare a coherent excited state we treat $\lambda(\tau)$ finite and non-perturbative, where it is inhomogeneous and periodic on the Euclidean time circle, $\lambda(\tau) = \lambda(\tau + \beta)$.
Such a deformation breaks Euclidean time translations, and consequently the Euclidean part of the path integral prepares a state at $t=0$ which is not in thermal equilibrium\footnote{In general, deformations by $O_3$ completely break supersymmetry and also break the $SU(4)_R$ R-symmetry group down to $SU(3)$ \cite{Distler:1998gb}.}, described a density operator $\hat{\rho}(0)$ with transition amplitudes
\bea
\small{\langle \chi_2|\hat\rho(0)| \chi_1\rangle = \int_{\substack{\tiny{\chi(\tau = \beta)=\chi_2}\\\tiny{\chi(\tau = 0)=\chi_1}}} {\cal D}\chi \exp\left(- S_{\text{E}} - \int_{{\cal C}_E} d\tau\,d^3x\lambda(\tau) O_3(\tau,x) \right)}.
\eea
In the Lorentzian segments, the sources $J_{1,2}$ are treated only perturbatively, so as to generate real-time response functions in this excited state. In particular we compute the following one-point function,
\be\label{onepoint}
\left<O_3\right>(t) = i\frac{\delta Z_{\text{CFT}}[\lambda, J_1, J_2]}{\delta J_1(t)}\bigg|_{J_1 = J_2 = 0} = \text{Tr}\left(\hat\rho(t) O_3\right)\,, 
\ee
where $\rho$ evolves in time according to the time-dependent Schrodinger equation giving rise to $\hat \rho(t) = U(t)\hat\rho(0)U^\dagger(t)$, with $U(t) = e^{-i H\,t}$ corresponding to the sourceless Lorentzian segment of the contour. 

Holographically, the deformed CFT path integral \eqref{pathint} is computed through the holographic dictionary $Z_{\text{CFT}}[\lambda_{\cal C}] = Z_{\text{bulk}}[\phi_{(1)} = \lambda_{\cal C}]$ where CFT sources $\lambda_{\cal C}$ give boundary conditions for a bulk scalar. In particular, the holographic dual of the $O_3$ operator is an $m^2L^2 = -3$ scalar, $\phi$, coupled to gravity in AdS$_5$. 
This is captured by the following consistent truncation of ${\cal N} = 8$ gauged supergravity \cite{Distler:1998gb, Gunaydin:1985cu, Khavaev:1998fb, Girardello:1999bd} which retains only this scalar field,\footnote{After appropriate field redefinitions by rescaling.}
\bea
S_{\text{bulk}} &=& \int d^5x \sqrt{|g|} \left(R - \frac{1}{2}(\partial \phi)^2 - V(\phi)\right), \label{Sbulk}\\
&& V(\phi) = \frac{3}{L^2} \cosh^2\left(\frac{\phi}{2}\right)(\cosh(\phi) - 5).
\eea
To describe the excited state, we seek solutions of the bulk equations of motion of $S_{\text{bulk}}$\footnote{Subject to suitable renormalisation \cite{Skenderis:2002wp}.} subject to the boundary conditions that the boundary metric is in the conformal class of the Schwinger-Keldysh contour ${\cal C}$, and with Dirichlet boundary conditions for the scalar $\phi$ in accordance with the holographic dictionary. Near the AdS boundary in Fefferman-Graham coordinates,
\be
\phi = z^{1} \phi_{(1)} + \ldots + z^{3}\phi_{(3)} + \ldots \label{scalarFG}
\ee
we have $\phi_{(1)} = \lambda(\tau)$ on the Euclidean segment of the contour and $\phi_{(1)} = 0$ otherwise. 
This saddle is obtained by regular, piecewise solutions to the bulk classical equations, glued together such that they exhibit continuity of field and first derivatives in the complex $t$ plane \cite{Skenderis:2008dh, Skenderis:2008dg}.  Since sources in ${\cal C}_{1,2}$ are treated perturbatively, the bulk Euclidean computation can be considered first, independently, as a way to generate initial data for the Lorentzian segments.\footnote{It is also worth noting that should equal sources be turned on in ${\cal C}_1$ and ${\cal C}_2$ the response would be causal and also not affect the Euclidean computation. When expanded perturbatively this generates fully-retarded correlation functions.}

\begin{figure}[h!]
\centering
\includegraphics[width=0.5\columnwidth]{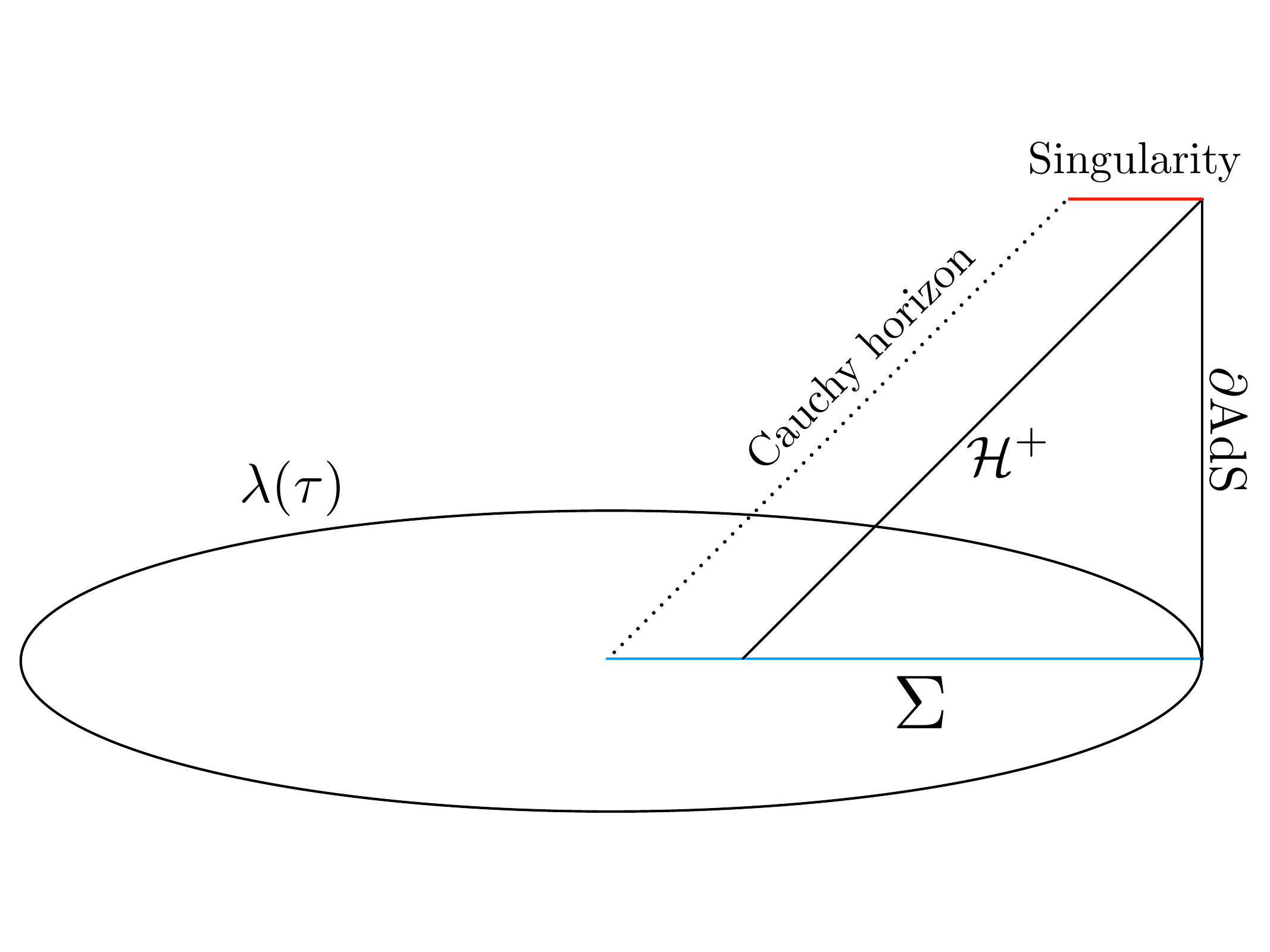}
\caption{An excited black hole state prepared using Euclidean path integral. The boundary of this spacetime is given by the closed time contour shown in figure \ref{fig:SK_contour}. The disk corresponds to a Euclidean geometry, deformed by sources for a relevant operator $\lambda(\tau)$ which break time translations and cause a departure from thermal equilibrium. The Lorentzian evolution of the state is given by the dynamical black hole spacetime indicated. Real-time correlators may be extracted by considering perturbations of this geometry.}
\label{fig:penrose}
\end{figure}

The Euclidean geometry that we construct resembles the classic cigar geometry, with the thermal circle smoothly shrinking to zero size in the interior. The geometry is deformed by the nontrivial boundary conditions, $\lambda(\tau)$, which backreacts in the bulk, deforming the cigar. With the Euclidean geometry constructed, taking the fields and their first derivatives on the $\tau = 0$ slice provides initial data for the Lorentzian evolution. We may take coordinates so that $\rho = 0$ is the tip of the cigar (where the $\tau$ circle shrinks to zero size), and parameterise the time circle $\tau(\varphi)$ using an angular coordinate $\varphi \sim \varphi + 2 \pi$ with $\tau(0) = 0$ and $\tau(2\pi) = \beta$ so that near the tip,
\be
ds^2 = C_2 \left(d\rho^2 + \rho^2 d\varphi^2\right) + C_3 dx_3^2,
\ee
with $C_2$, $C_3$ constant.
Taking $\tau = \varphi = 0$ as initial data for the Lorentzian, we see the tip of the cigar looks locally like Rindler. 
Without backreaction the causal development of this initial data is (a portion of) the exterior region of a black hole in an excited state. With backreaction, we will show that generically the event horizon of the black hole will intersect this surface at $\rho > 0$ and thus include a portion of interior black hole spacetime, as depicted in the causal diagram, figure \ref{fig:penrose}.\footnote{See also the Cauchy evolution for spacetimes describing expanding plasma \cite{Heller:2011ju, Heller:2012je}, exhibiting a similar causal structure.}

The rest of the paper proceeds as follows. 
In section \ref{sec:ansatz} we detail the ansatz and equations of motion to be used in the construction of the Euclidean and Lorentzian geometries.
In section \ref{sec:euclidean} we compute the Euclidean saddle corresponding to the deformed cigar, as a boundary-value problem, and read off initial data.
In section \ref{sec:lorentzian} we perform Cauchy time evolution of the initial data and monitor the approach to thermal equilibrium via a one-point function.
We discuss the maximal extension of the Lorentzian spacetime and the genericness of causal shadows in section \ref{sec:causality}.
We finish with a discussion in section \ref{sec:discussion}.

\section{Ansatz and equations of motion}\label{sec:ansatz}
We begin with constructing an ansatz for bulk fields in Lorentzian signature, and then describe how the ansatz is adjusted for the Euclidean segments. To motivate our choice, we start with the Schwarzschild-AdS$_5$ black brane written in Schwarzschild coordinates, with AdS radius $L=1$
\be
ds^2 = -\left(r^2 - \frac{r_H^4}{r^2}\right)dt^2 + \frac{dr^2}{\left(r^2 - \frac{r_H^4}{r^2}\right)} + r^2 dx_3^2.
\ee
We define a new radial coordinate via $r = r_H \sec\left(\frac{\pi \rho}{2}\right)$, so that $\rho = 0$ is the horizon and $\rho = 1$ is the AdS boundary,
\be
ds^2 = -\frac{3+\cos(\pi \rho)}{2}\tan^2\left(\frac{\pi \rho}{2}\right)r_H^2dt^2 + \frac{\pi^2 \sec^2\left(\frac{\pi \rho}{2}\right)}{6 + 2\cos(\pi \rho)}d\rho^2 + r_H^2 \sec^2\left(\frac{\pi \rho}{2}\right) dx_3^2.
\ee
We then pick $r_H=1/2$ for convenience. The metric is deformed from Schwarzschild by backreaction, which we allow by introducing the functions $A_L(t,\rho), B_L(t,\rho)$, giving our final ansatz,
\be
ds^2 = -\frac{3+\cos(\pi \rho)}{8}\tan^2\left(\frac{\pi \rho}{2}\right) A_L e^{-2B_L}dt^2 + \frac{\pi^2 \sec^2\left(\frac{\pi \rho}{2}\right)}{6 + 2\cos(\pi \rho)}\frac{d\rho^2}{A_L} + \frac{\sec^2\left(\frac{\pi \rho}{2}\right)}{4} dx_3^2, \label{metricansatz}
\ee
such that when $A_L = 1$, $B_L = 0$ we have planar Schwarzschild-AdS$_5$.
This is a particular choice of ADM metric with a lapse function but no shift. The scalar field is treated in first-order form, defining
\be
\Phi_L = (1-\rho^2)^{-1}\phi, \qquad \Pi_L = \frac{e^{B_L}}{A_L}(1-\rho^2)^{-1}\dot{\phi} \label{scalaransatz}
\ee
where dots denote $t$ derivatives.

The wave equation for the scalar becomes,
\bea
\dot{\Phi}_L &=& A_L e^{-B_L} \Pi_L,\label{eq:wave1}\\
\dot{\Pi}_L &=& \frac{13 \sin(\pi \rho) + 8 \sin(2\pi \rho) + \sin(3\pi \rho)}{64\pi^2 (1-\rho^2)}\times\nonumber\\
&&\qquad\qquad\partial_\rho\left((3+\cos(\pi \rho))\sec^2\left(\frac{\pi\rho}{2}\right)\tan\left(\frac{\pi\rho}{2}\right) A_L e^{-B_L} \partial_\rho((1-\rho^2)\Phi_L)\right)\nonumber\\
&&-\frac{(3+\cos(\pi \rho))\tan^2\left(\frac{\pi\rho}{2}\right)}{8(1-\rho^2)} e^{-B} V'\left[(1-\rho^2)\Phi_L\right].
\label{eq:wave2}
\eea
The Einstein equations $E^\mu_{\,\nu} = 0$ give the Hamiltonian constraint (from $E^t_{\, t}=0$)
\bea
A_L' &=&  - \left(\frac{8\pi \csc(\pi \rho)}{3+\cos(\pi \rho)} + \frac{8\pi \cot^2\left(\frac{\pi \rho}{2}\right)\csc(\pi \rho)}{3(3+\cos(\pi \rho))^2} (1-\rho^2)^2\Pi_L^2 + \frac{\cot\left(\frac{\pi \rho}{2}\right)}{3\pi} \left(\partial_\rho\left((1-\rho^2)\Phi_L\right)\right)^2\right)A_L\nonumber\\
&& - \frac{2\pi \csc(\pi \rho)}{3(3+\cos(\pi \rho))}V\left[(1-\rho)^2\Phi_L\right],\label{eq:ham}
\eea
the momentum constraint (from $E^t_{\,\rho}=0$)
\be
\dot{A}_L = -\frac{2}{3\pi} A_L^2 e^{-B_L} \cot\left(\frac{\pi \rho}{2}\right) (1-\rho^2)\Pi_L \partial_\rho\left((1-\rho^2)\Phi_L\right), \label{eq:mom}
\ee
as well as a slice condition (from $E^\rho_{\,\rho}=0$)
\be
B_L' = -\frac{2\pi \csc^4\left(\frac{\pi \rho}{2}\right)\sin(\pi \rho)}{3(3+\cos(\pi \rho))^2} \Pi_L^2 - \frac{\cot\left(\frac{\pi \rho}{2}\right)}{3\pi}\left(\partial_\rho\left((1-\rho^2)\Phi_L\right)\right)^2. \label{eq:slice}
\ee

\subsection{Fields and equations on ${\cal C}_E$}
The ans\"atze \eqref{metricansatz}, \eqref{scalaransatz} and the equations of motion \eqref{eq:wave1}, \eqref{eq:wave2}, \eqref{eq:ham}, \eqref{eq:mom}, \eqref{eq:slice} all apply to the Euclidean problem also, once we make the following replacements, 
\be
t = -i \varphi, \qquad A_L \to A_E, \quad B_L \to B_E, \quad \Phi_L \to \Phi_E, \quad \Pi_L \to i \Pi_E\,, \label{eq:euclideanmap}
\ee
where  $\varphi$ is a Euclidean time coordinate with period $2\pi$, and should be thought of as the angle coordinate in our numerical domain. 
For clarity, this gives $\Pi_E = \frac{e^{B_E}}{A_E}(1-\rho^2)^{-1}\dot{\phi}$ where the dot now denotes the $\varphi$ derivative.

\subsection{Fields and equations on ${\cal C}_2$}
The ans\"atze \eqref{metricansatz}, \eqref{scalaransatz} and the equations of motion \eqref{eq:wave1}, \eqref{eq:wave2}, \eqref{eq:ham}, \eqref{eq:mom}, \eqref{eq:slice} all apply to the Lorentzian problem on ${\cal C}_2$ also, once we make the following replacements, 
\be
t = -t_2, \qquad  A_L \to A_{L2}, \quad B_L \to B_{L2}, \quad\Phi_L \to \Phi_{L2}, \quad \Pi_L \to - \Pi_{L2}\,, \label{eq:L2map}
\ee
For clarity, this gives $\Pi_{L2} = \frac{e^{B_{L2}}}{A_{L2}}(1-\rho^2)^{-1}\dot{\phi}$ where the dot now denotes the $t_2$ derivative. 
In practise we do not need to perform an additional integration to determine the fields on ${\cal C}_2$ once they are known on ${\cal C}_1$, since we have no sources in the Lorentzian. Thus the solution on ${\cal C}_2$ is given immediately by
\be
A_{L2}(t_2) = A_{L}(-t_2), \quad B_{L2}(t_2) = B_{L}(-t_2), \quad \Phi_{L2}(t_2) = \Phi_{L}(-t_2), \quad \Pi_{L2}(t_2) = -\Pi_{L}(-t_2). \label{L2generation}
\ee
where we suppressed $\rho$ and as a reminder $t_2 \in [-t_f, 0]$ for some arbitrary final time $t_f$.

\subsection{Matching conditions} \label{sec:matching}
To construct the piecewise solution on ${\cal C}$ requires appropriate matching conditions between each of the three segments. These conditions guarantee $C^1$ of the fields in the complex $t$ plane, as required for the piecewise solution to be a saddle point.

Between ${\cal C}_E$ and ${\cal C}_1$ at $(\varphi, t_1) = (2\pi, 0)$ the solutions we study are time-reversal symmetric, so that $\Pi_E(2\pi,\rho) = \Pi_L(0,\rho) = \dot{A}_E(2\pi,\rho) = \dot{A}_L(0,\rho) = \dot{B}_E(2\pi,\rho) = \dot{B}_L(0,\rho) = 0$. In addition we require continuity of field values,
\bea
\Phi_E(2\pi,\rho) &=& \Phi_L(0,\rho),\\
A_E(2\pi,\rho) &=& A_L(0,\rho),\\
B_E(2\pi,\rho) &=& B_L(0,\rho),
\eea
which we enforce by using the Euclidean solution at $\varphi = 2\pi$ to specify the Lorentzian initial data at $t = t_1 = 0$.\footnote{Note that \emph{if} one requires the existence of a Fefferman-Graham expansion at this corner, $\lambda$ must be smooth since coefficients of the near-boundary expansion involves higher derivatives of $\lambda$ \cite{Skenderis:2008dg, Skenderis:2008dh}. Our choice of $\lambda$ is not smooth at the corner.}

Between ${\cal C}_E$ and ${\cal C}_2$ at $(\varphi, t_2) = (0,0)$ the matching conditions are similarly satisfied, again, all time derivatives are zero and 
\bea
\Phi_E(0,\rho) &=& \Phi_{L2}(0,\rho),\\
A_E(0,\rho) &=& A_{L2}(0,\rho),\\
B_E(0,\rho) &=& B_{L2}(0,\rho),
\eea
which are satisfied once the matching conditions between ${\cal C}_E$ and ${\cal C}_1$ are satisfied if the solution on ${\cal C}_2$ is generated by the map \eqref{L2generation}.

Between ${\cal C}_1$ and ${\cal C}_2$ at $(t_1, t_2) = (t_f, -t_f)$, we have the following conditions
\bea
\Phi_L(t_f,\rho) &=& \Phi_{L2}(-t_f,\rho),\\
A_L(t_f,\rho) &=& A_{L2}(-t_f,\rho),\\
B_L(t_f,\rho) &=& B_{L2}(-t_f,\rho),\\
\Pi_L(t_f,\rho) &=& -\Pi_{L2}(-t_f,\rho).
\eea
Which are all automatically satisfied for all $t_f > 0$ if the solution on ${\cal C}_2$ is generated by the map \eqref{L2generation}.

\section{Euclidean: The deformed cigar}\label{sec:euclidean}
In this section we construct regular Euclidean solutions corresponding to nontrivial source $\lambda$ for the $O_3$ operator.
The proper time at the boundary is $\tau$, with period $\beta$, we have that $\lambda(\tau) = \lambda(\tau + \beta)$ and $\lambda(0) = \partial_{\tau}\lambda(0) = 0$. We refer to the resulting geometry as the deformed cigar. 

In the numerical construction we work with the angular coordinate $\varphi$ with period $2\pi$, rather than $\tau$ directly.
Near the boundary, $\rho = 1$, we have the following behaviour, 
\bea
\Phi_E &=& \lambda(\tau(\varphi)) + O(1-\rho),\\
\Pi_E &=& \partial_{\tau}\lambda(\tau(\varphi))+ O(1-\rho),\\
A_E &=& 1 + O(1-\rho)^2,\\
B_E &=& B_\infty(\varphi) + O(1-\rho)^2.
\eea
Note that $B_\infty(\varphi)$ allows for boundary time reparameterisations, encoding the map from $\varphi$ to proper time $\tau$ through boundary.
The proper time at the boundary is given by,
\be
\tau(\varphi) = \int_0^\varphi e^{-B_\infty(\varphi')}d\varphi',
\ee
and the proper period of the Euclidean time circle is $\beta = \tau(2\pi)$.

In the interior near the tip of the cigar, $\rho = 0$, we have the following behaviour,
\bea
\Phi_E &=& \Phi_c + O(\rho)^2\\
\Pi_E &=& 0 + O(\rho)^2\\
A_E &=& -\frac{V(\Phi_c)}{12}+ O(\rho)^2\\
B_E &=& \log\left(-\frac{V(\Phi_c)}{12}\right)+ O(\rho)^2
\eea
where $\Phi_c$ is a constant.  This form of the metric means that near $\rho = 0$ we have
\be
ds^2 = \frac{3\pi^2}{2 \left(-V(\Phi_c)\right)} \left(\rho^2 d\varphi^2 +  d\rho^2\right) + \frac{1}{4} dx_3^2 + O(\rho)^2
\ee
so that $\varphi \sim \varphi + 2\pi$ is a good angular coordinate for a regular origin and there is no conical singularity. Roughly speaking, each term in the $\rho$ expansion corresponds to the contributions of higher multipoles on the thermal circle $\sim \rho^n e^{i n \varphi}$. In this work we specialise to solutions which only include the even multipoles, respecting a discrete Z$_2$ symmetry present in our choice of source function $\lambda(\tau)$.

\begin{figure}[h!]
\centering
\includegraphics[width=0.8\columnwidth]{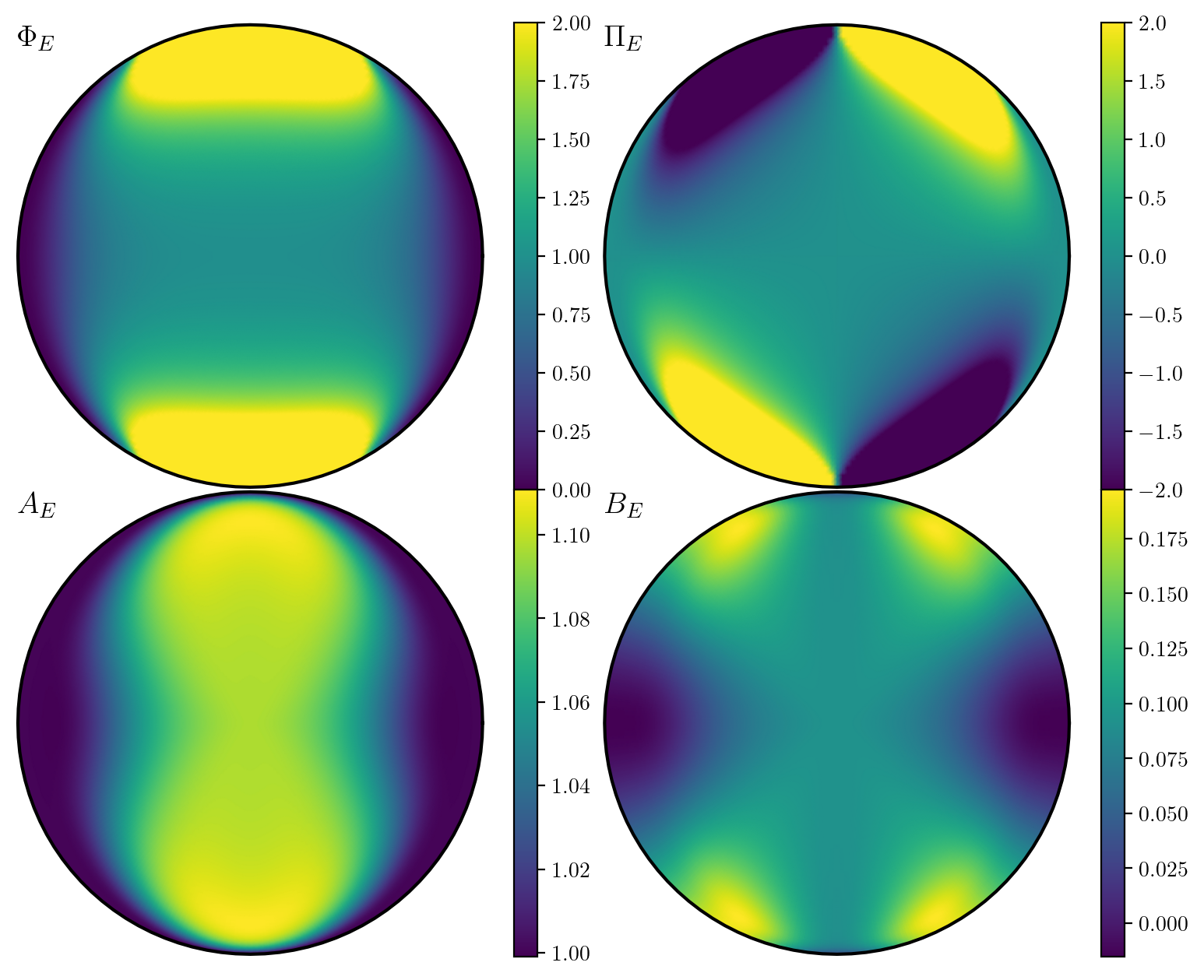}
\caption{The deformed cigar saddle point, forming the Euclidean part ${\cal C}_E$ of the bulk Schwinger-Keldysh contour ${\cal C}$ of figure \ref{fig:SK_contour}. These are polar plots with $\rho$ as the radial coordinate and $\varphi$ the angle, with $\rho = 1$ the edge of the disk and the conformal boundary of AdS. The surface $\tau = \varphi = 0$ is the horizontal interval starting at the centre and ending on the boundary on the right hand side in each plot. For clarity, the colour scales for $\Phi_E$ and $\Pi_E$ have been restricted to $[0,2]$ and $[-2,2]$ from their full ranges of $[0,21.8]$ and $[-65.4,65.4]$ respectively.}
\label{fig:euclidean_fields}
\end{figure}

\begin{figure}[h!]
\centering
\includegraphics[width=0.5\columnwidth]{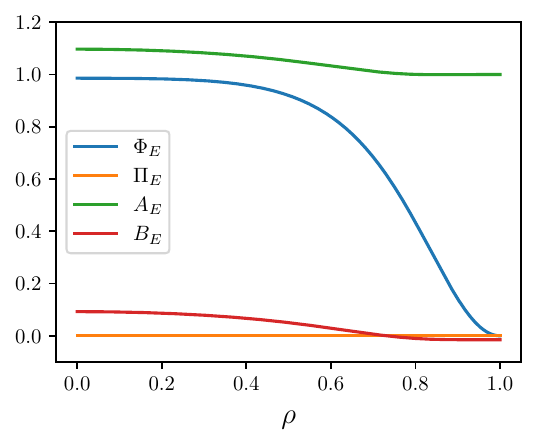}
\caption{A $\tau = 0$ ($\varphi = 0$) slice of the Euclidean path-integral prepared state shown in figure \ref{fig:euclidean_fields}. This serves as time-symmetric initial data for the Cauchy evolution of the Lorentzian fields $\Phi_L$, $\Pi_L$, $A_L$, $B_L$ through the matching conditions outlined in section \ref{sec:matching}. }
\label{fig:initial_data}
\end{figure}

We numerically solve \eqref{eq:wave1}, \eqref{eq:wave2} \eqref{eq:ham}, \eqref{eq:slice} (after applying the map \eqref{eq:euclideanmap}) as a boundary value problem, subject to the following regularity and boundary conditions, 
\begin{align}
\partial_\rho\Phi_E(\varphi, 0) &= 0 & \Phi_E(\varphi, 1) &= \lambda(\tau(\varphi))\\
\Pi_E(\varphi, 0) &= 0, &  \Pi_E(\varphi, 1) &= \partial_{\tau}\lambda(\tau(\varphi))\\
\partial_\rho A_E(\varphi, 0) &= 0, & A_E(\varphi, 1) &= 1\\
B_E(\varphi, 0) &= \log\left(A_E(\varphi,0)\right), & \partial_\rho B_E(\varphi, 1) &= 0
\end{align}
where the conditions near $\rho = 0$ are present to ensure regularity.
We use fourth-order finite differences for $\rho$ and Fourier spectral methods for $\varphi$, with the resulting nonlinear equations solved iteratively using the Newton-Raphson method. The momentum constraint \eqref{eq:mom}, is used to check the solution and perform continuum convergence tests, which are presented in appendix \ref{sec:convergence}. 
The solution for $\lambda = A\left(\frac{2\pi}{\beta}\right) \sin^{20}\left(\frac{2\pi}{\beta} \tau\right)$ is shown in figure \ref{fig:euclidean_fields}. Since $\lambda$ is peaked near $\tau = \beta/4$ and $\tau = 3\beta/4$ and $\Phi_E$ decays rapidly into the bulk, we used an amplitude $A = 20$ in order to generate appreciable backreaction along $\tau = \varphi = 0$. The resulting initial data for the Lorentzian evolution in section \ref{sec:lorentzian} is extracted at $\tau = \varphi = 0$ and is shown in figure \ref{fig:initial_data}.

\clearpage
\section{Lorentzian evolution}\label{sec:lorentzian}
In this section we study the Lorentzian segment of the Schwinger-Keldysh contour, without sources. This corresponds to a Cauchy development of the initial data extracted from the Euclidean segment of section \ref{sec:euclidean}, as shown in figure \ref{fig:initial_data}, via the matching conditions of section \ref{sec:matching} for time-reversal symmetric data at $\tau = 0$. We compute the bulk geometry and analyse the ringdown of one-point functions.

The Cauchy evolution scheme is as follows. Given $\Phi_L, \Pi_L, A_L, B_L$ at some time $t$, we use the scalar equation \eqref{eq:wave1}, \eqref{eq:wave2} to compute $\Phi_L$ and $\Pi_L$ on the next time slice at $t+\Delta t$. We then use the Hamiltonian constraint \eqref{eq:ham} and the slice condition \eqref{eq:slice}  to determine $A_L,B_L$ at $t+\Delta t$ by integrating along the slice. The momentum constraint \eqref{eq:mom} is not solved directly and is used to monitor the accuracy of the solution. We use second-order finite differences for $\rho$ derivatives, and fourth-order Runge-Kutta for time stepping. Boundary conditions at the AdS boundary follow from the required lack of CFT sources, i.e. $\Phi_L(t, 1) = 0$, $\Pi_L(t, 1) = 0$, $A_L(t,1) = 1$, and preserving the asymptotic time coordinate, $B_L(t,1) = B_\infty$ where $B_\infty$ is a constant read off from the initial data. At $\rho = 0$, the dynamics are frozen since the lapse function is zero there, consequently each Cauchy slice intersects the $\rho = 0$ corner.\footnote{In detail, \eqref{eq:wave2} implies that $\partial_t\Pi_L(t, 0) = 0$. Since the initial data gives $\Pi_L(0, 0) = 0$ we have $\Pi_L(t, 0) = 0$, which together with \eqref{eq:wave1} implies that $\partial_t\Phi_L(t, 0) = 0$. The initial data also satisfies $\partial_\rho\Pi_L(0, 0) = \partial_\rho\Phi_L(0, 0) = 0$, then subleading pieces of \eqref{eq:wave1}  and \eqref{eq:wave2} imply that $\partial_\rho\Pi_L(t, 0) = \partial_\rho\Phi_L(t, 0) = 0$. This in turn, through \eqref{eq:mom}, implies that $\partial_t A_L(t,0) = 0$.}

\begin{figure}[h!]
\centering
\includegraphics[width=0.8\columnwidth]{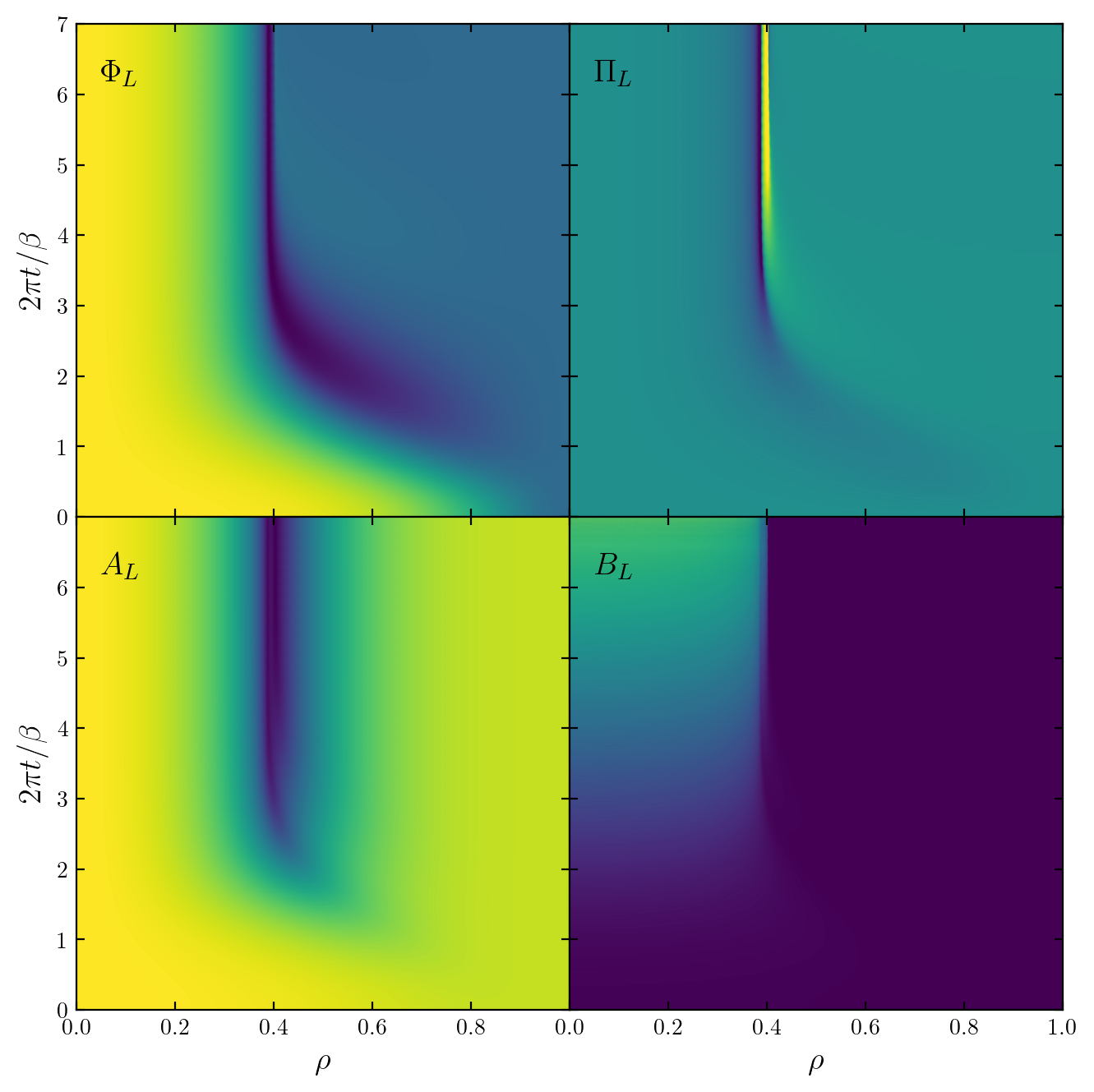}
\caption{Lorentzian Cauchy evolution giving the bulk dual of the segments ${\cal C}_1$, ${\cal C}_2$ of the Schwinger-Keldysh contour of figure \ref{fig:SK_contour}, developing from the deformed cigar solution of the Euclidean path integral of section \ref{sec:euclidean} at spatial resolution $N=2048$.}
\label{fig:lorentzian_fields}
\end{figure}

\begin{figure}[h!]
\centering
\includegraphics[width=0.95\columnwidth]{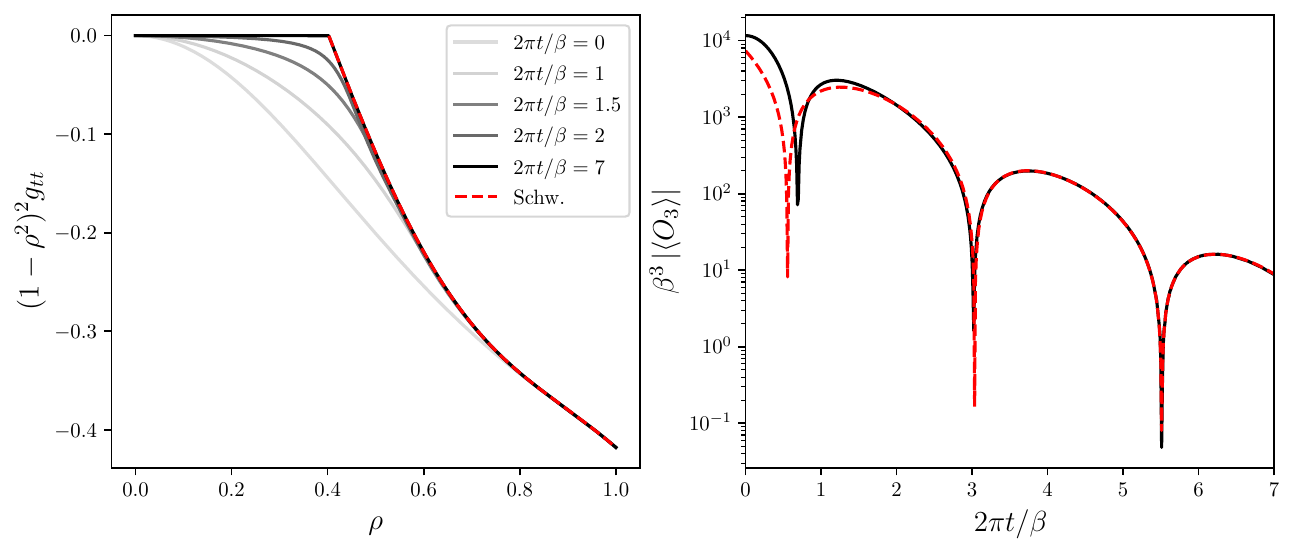}
\caption{Approach to thermal equilibrium at late times. \textbf{Left:} The metric component $g_{tt}$ (grey) converges to the known Schwarzschild solution \eqref{late_schwarzschild} outside the event horizon (red dashed). \textbf{Right:} The one-point function of the gaugino bilinear $O_3$ rings down to zero (black), as governed by the longest lived QNM of the late time Schwarzschild black hole (red dashed).}
\label{fig:lorentzian_late}
\end{figure}

The results of performing the Cauchy evolution is shown in figure \ref{fig:lorentzian_fields}. Convergence tests are given in Appendix \ref{sec:convergence}. 
We note that a portion of the late time dynamics are those of a Schwarzschild black brane whose event horizon lies inside the domain of development of the initial data. The Schwarzschild solution in these coordinates is given by
\be
A_L = \left(\frac{2\pi}{\tilde{\beta}}\right)^4 + 2\left(1-\left(\frac{2\pi}{\tilde{\beta}}\right)^4\right) \frac{\csc^2\left(\frac{\pi \rho}{2}\right)}{3+ \cos(\pi \rho)}, \qquad B_L = B_\infty, \qquad \Phi_L = \Pi_L = 0, \label{late_schwarzschild}
\ee
parameterised by the temperature $\tilde{\beta}^{-1}$. Note that $\tilde{\beta}$ should be distinguished from $\beta$, the (proper) period of the Euclidean circle from section \ref{sec:euclidean}.
This is indicated in figure \ref{fig:lorentzian_late} showing how $g_{tt}$ assumes the Schwarzschild solution outside the event horizon at late times, with $\tilde{\beta}/\beta \simeq 0.881$.  Also shown in figure \ref{fig:lorentzian_late} is the ringdown of the vev of the $O_3$ operator, consistent with the longest-lived QNM of this emergent Schwarzschild solution.

\begin{figure}[h!]
\centering
\includegraphics[width=\columnwidth]{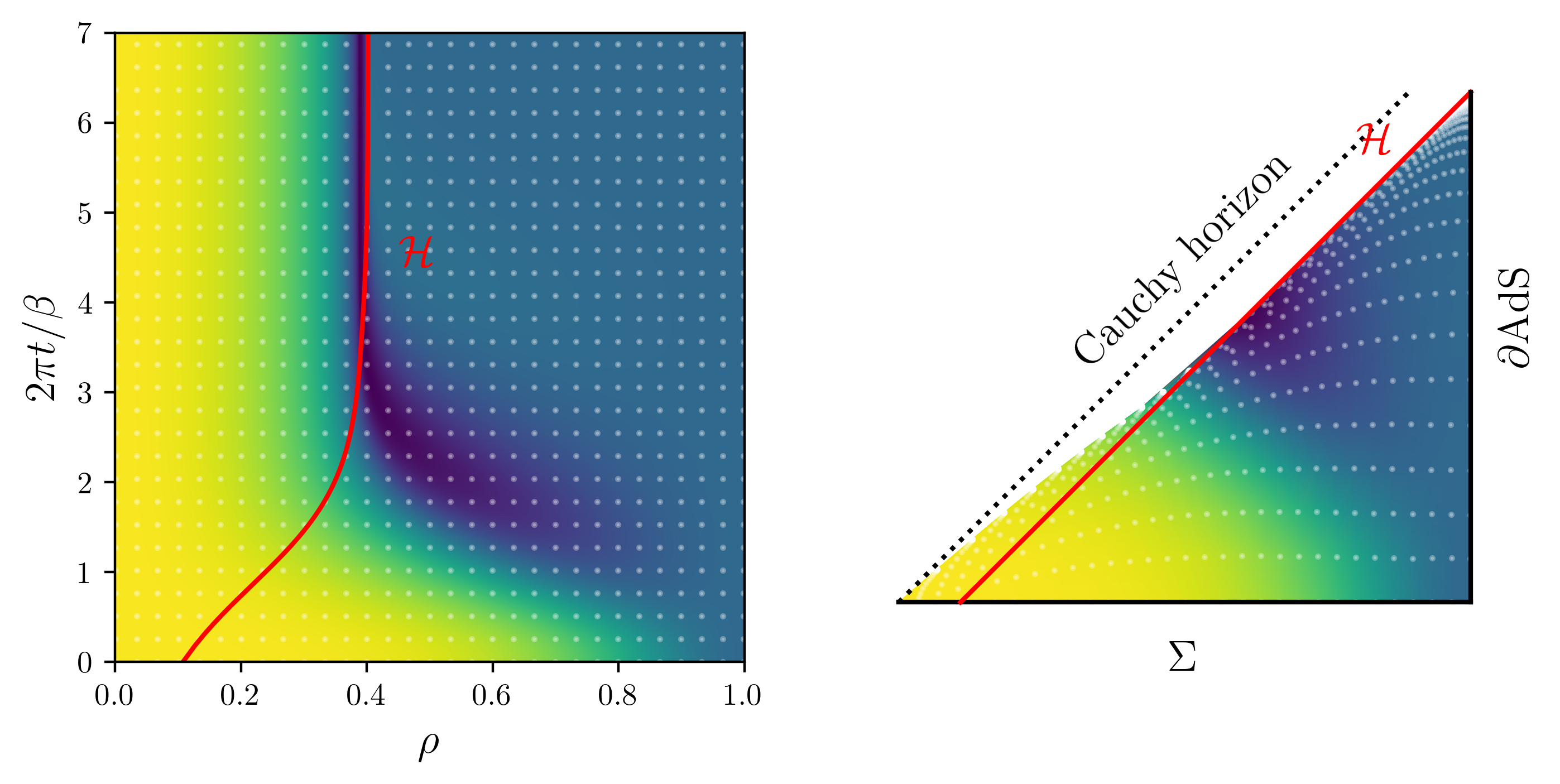}
\caption{Lorentzian part of the bulk Schwinger-Keldysh contour, with no sources turned on. A dynamical event horizon is present inside the domain of development at all times. The geometry outside the event horizon settles down to the Schwarzschild metric at late times, and the event horizon (red) intersects the initial data surface. Colour indicates the value of $\Phi_L$. \textbf{Left:} The numerical domain in coordinates $t, \rho$, with $\rho = 1$ the AdS boundary and $\rho = 0$ the tip of the deformed cigar in the initial data. \textbf{Right:} Causal diagram for the same simulation constructed by shooting pairs of null geodesics through the numerical domain. The white dots in both are a guide to the eye for the mapping from one to the other. The dashed line denotes the putative Cauchy horizon for data on $\Sigma$ (i.e. boundary of the domain of development). }
\label{fig:lorentzian_scalar}
\end{figure}

An analysis of the causal structure is shown in figure \ref{fig:lorentzian_scalar}. The event horizon was obtained by fitting to \eqref{late_schwarzschild} at late time to find the horizon radius, then shooting an outgoing null geodesic backwards in time from there. We obtained the conformal diagram by shooting both ingoing and outgoing null geodesics backwards in time from each indicated spacetime point to see where they intersected the initial data surface. Rays that hit the boundary were reflected and then integrated further until they also hit the initial data surface. This data is sufficient to directly compute the conformal map.

\begin{figure}[h!]
\centering
\includegraphics[width=\columnwidth]{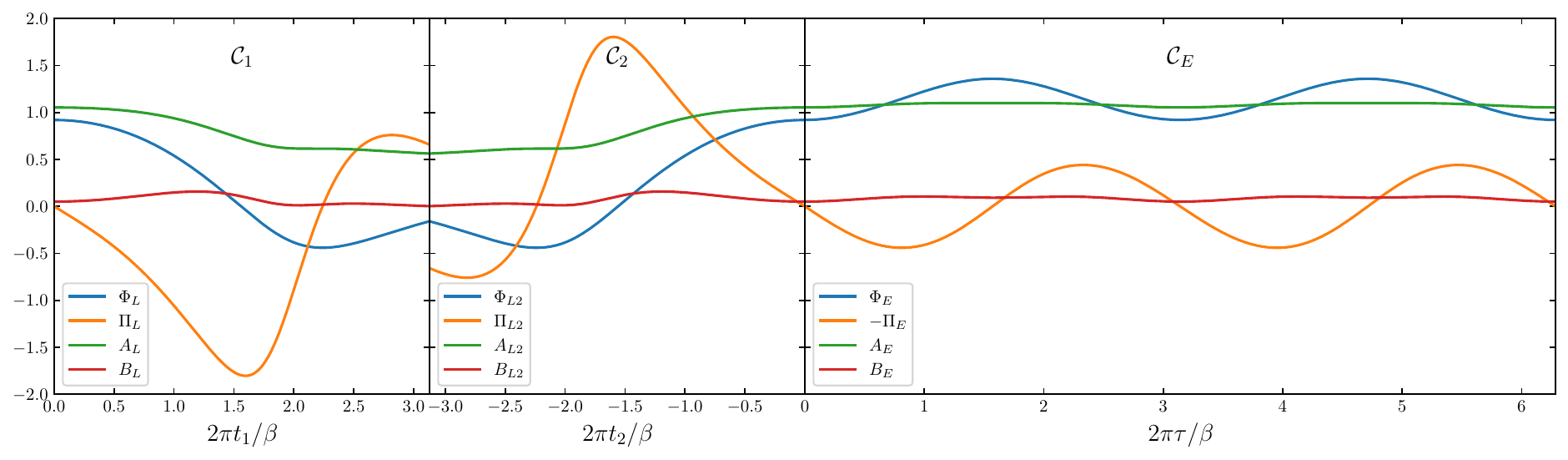}
\caption{Time evolution of the bulk fields at $\rho = 0.5$ in each segment of the bulk dual of the Schwinger-Keldysh contour of figure \ref{fig:SK_contour}, illustrating the matching of fields between each segment outlined in section \ref{sec:matching}. We have chosen an arbitrary choice of final time for the contour, $t_f = \beta/2$, despite knowing the bulk solution up to $t_f \simeq 1.1 \beta$, to highlight that matching holds at any choice of $t_f$, due to a lack of sources in the Lorentzian segments. In accordance with the conditions of section \ref{sec:matching}, the fields $\Phi, A, B$ match at the joins between any two segments, while $\Pi$ (orange curve) is zero at $(\tau, t_1) = (\beta, 0)$ and at $(\tau, t_2) = (0,0)$, and changes sign at $(t_1, t_2) = (t_f, -t_f)$ since $\partial_{t_1} \Phi_L = -\partial_{t_2} \Phi_{L2}$ at this join. In the rightmost plot we choose to show $-\Pi_E$ rather than $\Pi_E$ to further illustrate the (unnecessary) matching of the second time derivative of $\Phi$, i.e. $\partial_{t_1}^2 \Phi_L = -\partial_\tau^2 \Phi_E$ at $\tau = 0, \beta$.} 
\label{fig:bulkmatching}
\end{figure}

Finally we illustrate the matching of the bulk solution between each of the segments of ${\cal C}$ in figure \ref{fig:bulkmatching}, at an arbitrary choice of radius $\rho = 1/2$. We also picked an arbitrary final time $t_f = \beta/2$ to illustrate that the bulk matching between ${\cal C}_1$ and ${\cal C}_2$ holds for any choice of $t_f$, as discussed in section \ref{sec:matching}. This is a consequence of the lack of sources in these segments. We observe that the number of derivatives in the complex $t$ plane which are continuous at the corners $\tau = 0$ and $\tau = \beta$ is higher than strictly necessary for a solution of the Einstein equations. Here this is a consequence of choosing an order-20 zero in $\lambda(\tau)$ at $\tau = 0$.

\clearpage
\section{Causal shadows}\label{sec:causality}
\begin{figure}[h!]
\centering
\includegraphics[width=0.6\columnwidth]{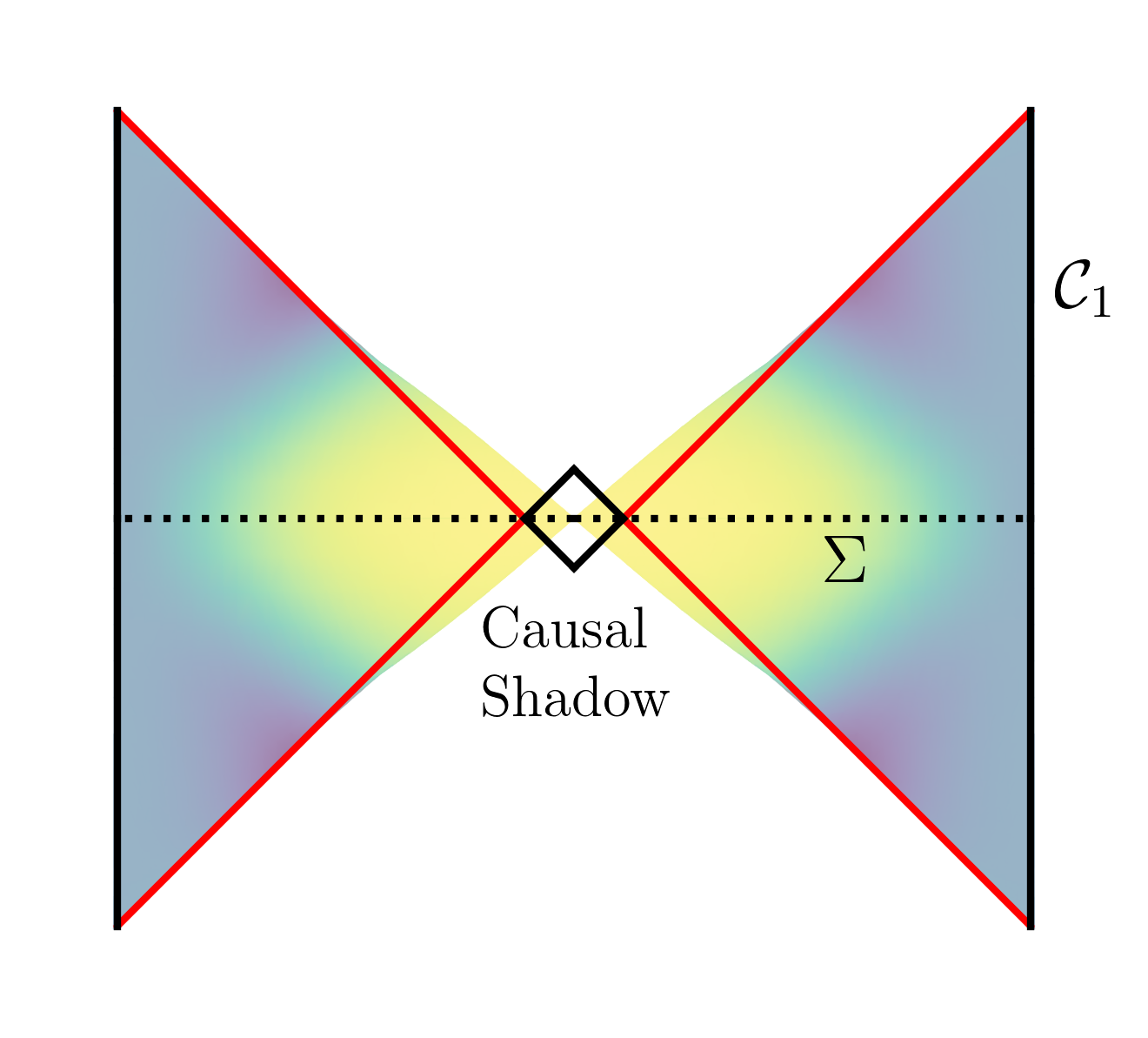}
\caption{The maximal extension of the Lorentzian segment ${\cal C}_1$ in the bulk, constructed directly from numerical solutions of the Einstein-scalar field system. Whilst far from thermal equilibrium, this extension exists  because of a combination of time-symmetric initial data, and a $t \to t - i \beta/2$ symmetry in its Euclidean preparation. The red lines are the event horizons associated to the right and left boundaries, the vertical black lines. The diagonal black lines enclose a region of spacetime spacelike separated from both boundaries, the `causal shadow'. The shading is the scalar field matter distribution that supports this black hole, shown only where it is known from numerical solutions. Black hole singularities are not indicated.}
\label{fig:maximal_extension}
\end{figure}

Since we preserved a Z$_2$ symmetry $\lambda(\tau) = \lambda(\tau + \beta/2)$ the dynamical black hole has an identical asymptotic region under $t \to t - i \beta/2$. The initial data is also time symmetric. These symmetries allow us to construct the maximal extension of the Lorentzian spacetime, by combining images under $t \to t - i \beta/2$ and $t\to -t$. The associated causal diagram is shown in figure \ref{fig:maximal_extension} and displays a `causal shadow' \cite{Headrick:2014cta}, a region of the bulk spacetime which is spacelike separated from both components of the boundary.\footnote{In \cite{Headrick:2014cta} causal shadows are defined more generally as bulk regions spacelike separated from ${\cal D}[{\cal A}]\cup {\cal D}[{\cal A}_c]$ where ${\cal A}$ is a spatial subregion and ${\cal D}[{\cal A}]$ its boundary domain of dependence. In the present case, ${\cal A}$ is an entire Cauchy slice of one component of the boundary.} Similar black hole causal structure was previously seen from symmetric collapse of null shells as described by Vaidya \cite{Headrick:2014cta}, for three dimensional Janus black holes \cite{Nakaguchi:2014eiu} and for three dimensional black holes with more than two asymptotic regions \cite{Skenderis:2009ju, AlBalushi:2020kso}. Indeed, \cite{Skenderis:2009ju} interpreted such three-dimensional wormholes within the in-in formalism using real-time holography. Finally, in a recent series of papers \cite{Balasubramanian:2023xyd,Balasubramanian:2022gmo,Balasubramanian:2022lnw}, it was argued that novel wormhole configurations exhibiting causal shadows can be used for counting black hole microstates, explaining the microscopic origin of the Bekenstein-Hawking black hole entropy.

\begin{figure}[h!]
\centering
\includegraphics[width=0.6\columnwidth]{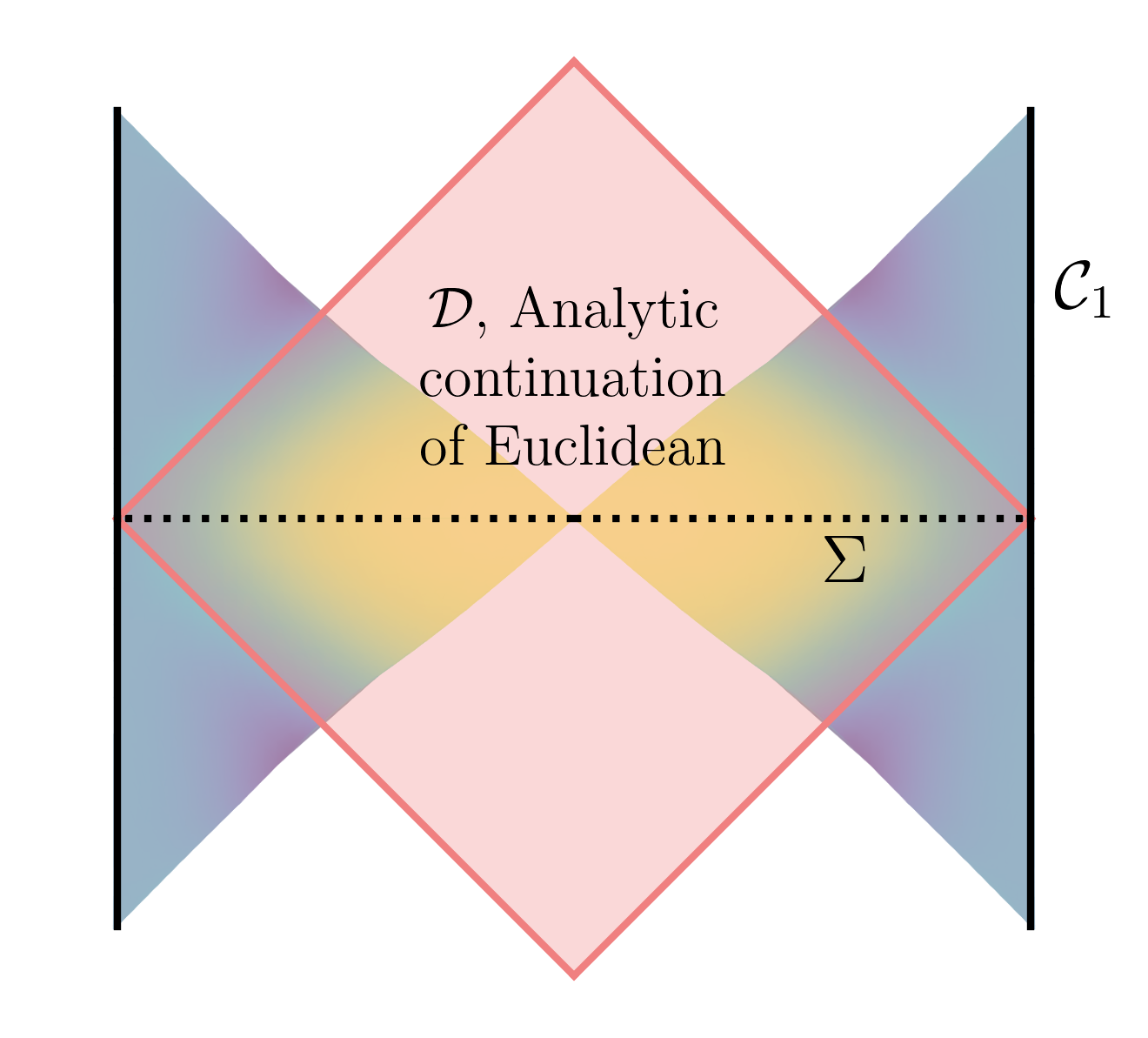}
\caption{There is a spacetime region ${\cal D}$ in the Lorentzian segment which is insensitive to the choice of time-symmetric Lorentzian boundary conditions, by causality. It follows that ${\cal D}$ depends only on the deformed Euclidean cigar solution. Provided the Euclidean source function $\lambda(\tau)$ is itself analytic, then the solution in ${\cal D}$ follows from analytic continuation of the Euclidean solution. Outside ${\cal D}$ the choice of Lorentzian sources become important.}
\label{fig:diamond}
\end{figure}

One may be concerned about the geometry lying beyond the Cauchy horizon for the evolution we have performed; does it exist? There is a simple argument that shows that it must. First, we note that the Lorentzian spacetime, globally, is not the analytic continuation of the Euclidean one under $\tau \to i t$. This is easy to see, since $\lambda(\tau)$ would continue to $\lambda(it)$ on the Lorentzian boundary, however the solutions we have constructed have zero sources there. However, the causal diamond ${\cal D}$ as the domain of dependence of the double-sided Cauchy initial data (formed from the union of $\tau = 0$ and $\tau = \beta/2$ Euclidean slices, shown as the dotted line in figure \ref{fig:maximal_extension}) is independent of the choice of source function in the Lorentzian. This is illustrated in figure \ref{fig:diamond}. This follows from causality; any choice of time-symmetric Lorentzian source function is unable to affect the region ${\cal D}$. The solution inside ${\cal D}$ is then the appropriate analytic continuation of the Euclidean solution, analogous to the continuation from two back-to-back Rindler wedges into the full Minkowski spacetime.  As a result, the regions immediately to the future and past of the tip of the cigar are guaranteed to exist, provided the Euclidean source choice $\lambda(\tau)$ is an analytic function. The causal shadow lies inside ${\cal D}$, though note however the size of the causal shadow is a consequence of global causal structure and thus cannot be determined from analytic continuation of the Euclidean alone.

It is natural to ask whether causal shadows are generic for CFT states prepared by Euclidean path integrals, deforming the thermal circle. Since the tip of the Euclidean cigar at $\rho = 0$ corresponds to a trapped surface in the Lorentzian, the event horizon ${\cal H}$ must be located at $\rho \geq 0$ for matter obeying null energy condition. For a causal shadow to exist one requires that the horizon grows.

In more detail, let $U$ be the tangent vector for a null geodesic congruence whose motion is restricted to the $t,\rho$ plane, i.e. $U$ only has nontrivial components in the $t,\rho$ directions. The null Raychaudhuri equation for geodesics in this congruence is given by
\be
\dot\theta = -\frac{1}{3}\theta^2 - \sigma^2 + \omega^2 - R_{\mu\nu}U^\mu U^\nu, \label{eq:ray1}
\ee
where dots denote derivatives with respect to the affine parameter $s$, with expansion $\theta$, and here vanishing shear and rotation $\sigma_{ab} = \omega_{ab} = 0$. Consider arbitrary matter contributions obeying the null energy condition, so that $T_{\mu\nu}U^\mu U^\nu \geq 0$, then
\be
\dot\theta = -\frac{1}{3}\theta^2 - \frac{1}{2}T_{\mu\nu}U^\mu U^\nu. \label{eq:ray2}
\ee
As usual, if $\theta < 0$ anywhere along a geodesic $\gamma$, then it will evolve to $-\infty$ at finite $s$, since 
\be
\dot\theta \leq -\frac{1}{3}\theta^2, \quad \theta(s_0) = \theta_0 \;\implies\; \theta \leq \frac{\theta_0}{1+ \frac{1}{3}(s - s_0)\theta_0},\label{eq:raysing}
\ee
and the right hand side goes to $-\infty$ at finite $s$. 

Now consider the geodesic $\gamma$ that starts at the tip of the Euclidean cigar, $\rho = t = 0$. This is the generator of the Cauchy horizon. One can show that at the tip, $\theta(0) = 0$. From \eqref{eq:ray2}, if $T_{\mu\nu}U^\mu U^\nu = 0$ along $\gamma$ then $\theta = 0$ everywhere along $\gamma$, and thus there is no sign of a singularity being encountered at finite $s$. Consider some matter falling into the Cauchy horizon such that $T_{\mu\nu}U^\mu U^\nu > 0$ there. The effect is that $\dot\theta < 0$ and hence $\theta$ becomes negative. Hence, a singularity is reached at finite $s$ and $\gamma$ must lie inside the event horizon ${\cal H}$. In summary, if $T_{\mu\nu}U^\mu U^\nu >0$ anywhere along the Cauchy horizon, then we necessarily have a causal shadow. 

The condition $T_{\mu\nu}U^\mu U^\nu >0$ will be obeyed generically by matter falling into the black hole. In the case of a scalar field, $T_{\mu\nu}U^\mu U^\nu = (U\cdot \nabla\phi)^2$. 
Recall that we can access a portion of the Lorentzian geometry by analytic continuation, into ${\cal D}$. We can use this to test the condition $T_{\mu\nu}U^\mu U^\nu > 0$ on the Cauchy horizon for a scalar field.
For example, in the neighbourhood of $\rho = 0$ we may describe the Euclidean geometry and the scalar field by
\be
ds^2 = d\rho^2 + \rho^2 d\tau^2, \qquad \phi(\tau, \rho) = \phi_0 + \phi_2\rho^2\cos(2\tau)
\ee
where we have specialised to $\beta = 2\pi$ for convenience, and suppressed spatial directions and additional corrections in $\rho$. Because we are in the neighbourhood of the tip of the cigar, the analytic continuation $\tau = i t$ produces the Rindler spacetime. Now the Cauchy horizon is just the Rindler horizon at $\rho = 0$, generated by $U = \partial_t$. The causal development of the $t=0$ data still only evolves into the exterior of Rindler, however we can perform a further analytic continuation from Rindler into Minkowski, ${\cal D}$, $-dT^2 + dX^2$, by $T = \rho \sinh(t), X = \rho\cosh(t)$, where $U = X\partial_T + T\partial_X$. The scalar field becomes
\be
\phi = \phi_0 + \phi_2 (X^2  + T^2) \implies T_{\mu\nu}U^\mu U^\nu = 16\phi_2 T^2 X^2,
\ee
which is non-zero on $T=X$.
Similar conclusions hold for higher multipoles of the scalar in the small $\rho$ neighbourhood, $\phi_n\rho^n e^{i n \tau}$.
Hence provided the Euclidean time translations are broken, so that the Euclidean quadrupole $\phi_2$ or higher moments $\phi_n$ are non-zero, we have $T_{\mu\nu}U^\mu U^\nu > 0$ on the Cauchy horizon and a causal shadow is formed.\footnote{Note that if Euclidean-time translations were not broken, the corresponding cigar geometry would describe an equilibrium state, whose Lorentzian geometry is obtained under analytic continuation and contain constant Lorentzian sources.}

\section{Discussion}\label{sec:discussion}
Real time holography is relatively unexplored compared to its Euclidean counterpart.
The real time formalism comes with distinct advantages, allowing the computation of real-time observables in field theory, such as all combinations of retarded and advanced correlation functions.
We applied real time holographic techniques to study excited thermal states in ${\cal N} = 4$ SYM. These were prepared by introducing relevant deformations with sources that break the translation symmetry of the thermal circle. The deformation led to non-perturbative, coherent departures from thermal equilibrium described by semi-classical gravity. We studied the approach to thermal equilibrium in real-time under Lorentzian evolution. This required the use of numerical relativity in order to construct the bulk dual.

We highlighted the connection between the geometry of dynamical black hole spacetimes and non-equilibrium field theory, summarised in the causal diagrams of figures \ref{fig:penrose} and \ref{fig:lorentzian_scalar}. The event horizon is always present, intersecting the initial data surface away from the tip of the deformed cigar geometry. We demonstrated that this feature is generic in the presence of matter obeying the null energy condition, if one breaks translations on the Euclidean time circle. It corresponds to a `causal shadow' in the maximal extension of the Lorentzian spacetime.

Note that because the deformed cigar is regular at $\rho = 0$, one cannot prepare a finite sum of QNMs with these techniques. This is because QNMs are not regular at $\rho = 0$ where they behave as an ingoing mode $\sim \rho^{-i \omega}e^{-i \omega t}$. Instead we can think of preparing an infinite sum of them. From this sum, at late times the longest lived QNM dominates, and hence the solution becomes increasingly singular in these coordinates (as figure \ref{fig:lorentzian_scalar} demonstrates this is purely a coordinate artefact). An explicit construction of regular Cauchy data from an infinite sum of QNMs was given for low dimensional examples in \cite{Bzowski:2022kgf}. 

There are several open questions that arise for the formalism itself. We focused on saddles which were constructed by piecewise gluing of Euclidean and Lorentzian segments. However, it is not obvious whether there are other saddle points whose metrics are complex in the interior and whether they should contribute to the gravitational path integral.\footnote{See \cite{Witten:2021nzp} for a recent discussion of complex metrics and their role in gravitational path integrals.} 
It would also be interesting to develop a numerical method to find bulk solutions when $J_a \equiv J_1 - J_2 \neq 0$ non-perturbatively.
The associated PDE problem is of mixed-type; with an appropriate treatment of gauge redundancy (for example using the De-Turck method \cite{Headrick:2009pv}), the Euclidean equations have an elliptic character, the Lorentzian equations have a hyperbolic character, and they must be compatible with one another along the gluing surfaces.\footnote{In contrast when $J_a = 0$ the response sources in the Lorentzian is causal, and can be treated as an initial-boundary-value problem with initial data read off from the Euclidean path integral, just as we did in this work.} 

More generally one may hope that the exploration of real-time holography may be fruitful in both directions of the duality; either for using gravity to gain new insight into dynamical phenomena in field theory \cite{Hohenberg:1977ym}, or for motivating the construction of new exotic black holes as solutions of mixed-signature boundary value problems, as we have done here.

\section*{Acknowledgements}
It is a pleasure to thank Adam Bzowski, Carsten Gundlach, Michal Heller, Romuald Janik, Pavel Kovtun and Toby Wiseman for discussions.
B.W. thanks KITP Santa Barbara for hospitality during the programme ``The Many Faces of Relativistic Fluid Dynamics''.
 C.P. acknowledges support from a Royal Society - Science Foundation Ireland University Research Fellowship via grant URF/R1/211027. 
B.W. is supported by a Royal Society University Research Fellowship and in part by the Science and Technology Facilities Council (Consolidated Grant `Exploring the Limits of the Standard Model and Beyond').
This research was supported in part by the National Science Foundation under Grant No. NSF PHY-1748958.

\appendix
\section{Convergence tests}
\label{sec:convergence}
\begin{figure}[h!]
\centering
\includegraphics[width=0.5\columnwidth]{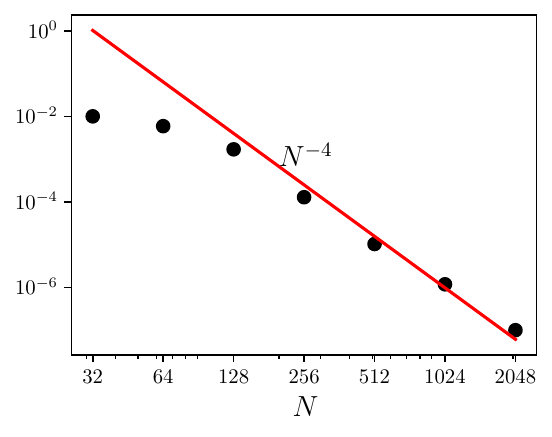}
\caption{Continuum convergence tests of the maximum violation of the momentum constraint \eqref{eq:mom}, for the Euclidean solutions of section \ref{sec:euclidean}. Fourth-order finite differences are used for $\rho$ with $N$ points, while for $t$ we use spectral methods with a fixed $128$ Fourier modes. The dominant truncation error is due to the finite $\rho$ resolution and the plot shows the correct error scaling towards zero as $N$ is increased.}
\label{fig:euclidean_convergence}
\end{figure}

\begin{figure}[h!]
\centering
\includegraphics[width=\columnwidth]{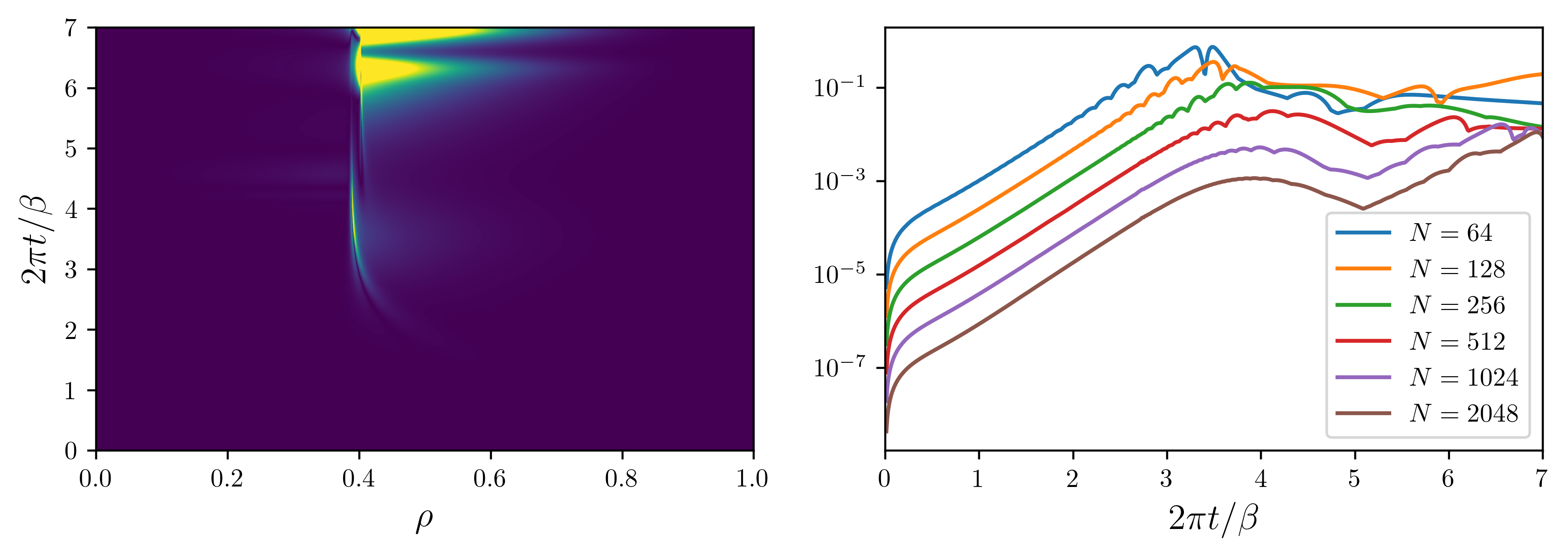}
\caption{Continuum convergence test for the Lorentzian evolution of section \ref{sec:lorentzian}. \textbf{Left:} Spacetime dependence of the momentum constraint  \eqref{eq:mom} at $N = 2048$ showing where the numerical solution breaks down as gradients become increasingly large in these coordinates at late times near the horizon. \textbf{Right:} Convergence of the maximum violation of the momentum constraint \eqref{eq:mom} on a given timeslice, as a function of time, for the Lorentzian evolution of section \ref{sec:lorentzian}. Time step is fixed at $\Delta t = 10^{-3}$ in RK4 and $N$ is the number of points in a second-order finite difference for spatial derivatives.}
\label{fig:lorentzian_constraint}
\end{figure}

To assess convergence towards the continuum solution we monitor the momentum constraint \eqref{eq:mom}, which is the only equation not solved directly in our numerical scheme. In the Euclidean calculation of section \ref{sec:euclidean}, we utilise fourth-order finite differences, and the appropriate rate of convergence to zero is exhibited in the solutions, as demonstrated in figure \ref{fig:euclidean_convergence}.
For the Lorentzian solutions of section \ref{sec:lorentzian}, we show convergence of the momentum constraint in figure \ref{fig:lorentzian_constraint}. The constraint violation grows with time as the black hole settles down, which can be attributed to large gradients forming in the solution just inside the event horizon. This scales to zero as the resolution is increased.

\bibliographystyle{ytphys}
\bibliography{refs}

\providecommand{\href}[2]{#2}\begingroup\raggedright\begin{thebibliography}{10}

\bibitem{Maldacena:1997re}
J.~M. Maldacena, ``{The Large N limit of superconformal field theories and
  supergravity},'' \href{http://dx.doi.org/10.4310/ATMP.1998.v2.n2.a1}{{\em
  Adv. Theor. Math. Phys.} {\bfseries 2} (1998) 231--252},
  \href{http://arxiv.org/abs/hep-th/9711200}{{\ttfamily arXiv:hep-th/9711200}}.

\bibitem{Skenderis:2008dh}
K.~Skenderis and B.~C. van Rees, ``{Real-time gauge/gravity duality},''
  \href{http://dx.doi.org/10.1103/PhysRevLett.101.081601}{{\em Phys. Rev.
  Lett.} {\bfseries 101} (2008) 081601},
  \href{http://arxiv.org/abs/0805.0150}{{\ttfamily arXiv:0805.0150 [hep-th]}}.

\bibitem{Skenderis:2008dg}
K.~Skenderis and B.~C. van Rees, ``{Real-time gauge/gravity duality:
  Prescription, Renormalization and Examples},''
  \href{http://dx.doi.org/10.1088/1126-6708/2009/05/085}{{\em JHEP} {\bfseries
  05} (2009) 085}, \href{http://arxiv.org/abs/0812.2909}{{\ttfamily
  arXiv:0812.2909 [hep-th]}}.

\bibitem{Glorioso:2018mmw}
P.~Glorioso, M.~Crossley, and H.~Liu, ``{A prescription for holographic
  Schwinger-Keldysh contour in non-equilibrium systems},''
  \href{http://arxiv.org/abs/1812.08785}{{\ttfamily arXiv:1812.08785
  [hep-th]}}.

\bibitem{Belin:2020zjb}
A.~Belin and B.~Withers, ``{From sources to initial data and back again: on
  bulk singularities in Euclidean AdS/CFT},''
  \href{http://dx.doi.org/10.1007/JHEP12(2020)185}{{\em JHEP} {\bfseries 12}
  (2020) 185}, \href{http://arxiv.org/abs/2007.10344}{{\ttfamily
  arXiv:2007.10344 [hep-th]}}.

\bibitem{Loganayagam:2022zmq}
R.~Loganayagam, M.~Rangamani, and J.~Virrueta, ``{Holographic open quantum
  systems: toy models and analytic properties of thermal correlators},''
  \href{http://dx.doi.org/10.1007/JHEP03(2023)153}{{\em JHEP} {\bfseries 03}
  (2023) 153}, \href{http://arxiv.org/abs/2211.07683}{{\ttfamily
  arXiv:2211.07683 [hep-th]}}.

\bibitem{Pantelidou:2022ftm}
C.~Pantelidou and B.~Withers, ``{Thermal three-point functions from holographic
  Schwinger-Keldysh contours},''
  \href{http://dx.doi.org/10.1007/JHEP04(2023)050}{{\em JHEP} {\bfseries 04}
  (2023) 050}, \href{http://arxiv.org/abs/2211.09140}{{\ttfamily
  arXiv:2211.09140 [hep-th]}}.

\bibitem{vanRees:2009rw}
B.~C. van Rees, ``{Real-time gauge/gravity duality and ingoing boundary
  conditions},''
  \href{http://dx.doi.org/10.1016/j.nuclphysbps.2009.07.078}{{\em Nucl. Phys. B
  Proc. Suppl.} {\bfseries 192-193} (2009) 193--196},
  \href{http://arxiv.org/abs/0902.4010}{{\ttfamily arXiv:0902.4010 [hep-th]}}.

\bibitem{Botta-Cantcheff:2015sav}
M.~Botta-Cantcheff, P.~Mart\'\i{}nez, and G.~A. Silva, ``{On excited states in
  real-time AdS/CFT},'' \href{http://dx.doi.org/10.1007/JHEP02(2016)171}{{\em
  JHEP} {\bfseries 02} (2016) 171},
  \href{http://arxiv.org/abs/1512.07850}{{\ttfamily arXiv:1512.07850
  [hep-th]}}.

\bibitem{Christodoulou:2016nej}
A.~Christodoulou and K.~Skenderis, ``{Holographic Construction of Excited CFT
  States},'' \href{http://dx.doi.org/10.1007/JHEP04(2016)096}{{\em JHEP}
  {\bfseries 04} (2016) 096}, \href{http://arxiv.org/abs/1602.02039}{{\ttfamily
  arXiv:1602.02039 [hep-th]}}.

\bibitem{Botta-Cantcheff:2017qir}
M.~Botta-Cantcheff, P.~J. Mart\'\i{}nez, and G.~A. Silva, ``{Interacting fields
  in real-time AdS/CFT},''
  \href{http://dx.doi.org/10.1007/JHEP03(2017)148}{{\em JHEP} {\bfseries 03}
  (2017) 148}, \href{http://arxiv.org/abs/1703.02384}{{\ttfamily
  arXiv:1703.02384 [hep-th]}}.

\bibitem{Marolf:2017kvq}
D.~Marolf, O.~Parrikar, C.~Rabideau, A.~Izadi~Rad, and M.~Van~Raamsdonk,
  ``{From Euclidean Sources to Lorentzian Spacetimes in Holographic Conformal
  Field Theories},'' \href{http://dx.doi.org/10.1007/JHEP06(2018)077}{{\em
  JHEP} {\bfseries 06} (2018) 077},
  \href{http://arxiv.org/abs/1709.10101}{{\ttfamily arXiv:1709.10101
  [hep-th]}}.

\bibitem{Botta-Cantcheff:2018brv}
M.~Botta-Cantcheff, P.~J. Mart\'\i{}nez, and G.~A. Silva, ``{The Gravity Dual
  of Real-Time CFT at Finite Temperature},''
  \href{http://dx.doi.org/10.1007/JHEP11(2018)129}{{\em JHEP} {\bfseries 11}
  (2018) 129}, \href{http://arxiv.org/abs/1808.10306}{{\ttfamily
  arXiv:1808.10306 [hep-th]}}.

\bibitem{Botta-Cantcheff:2019apr}
M.~Botta-Cantcheff, P.~J. Mart\'\i{}nez, and G.~A. Silva, ``{Holographic
  excited states in AdS Black Holes},''
  \href{http://dx.doi.org/10.1007/JHEP04(2019)028}{{\em JHEP} {\bfseries 04}
  (2019) 028}, \href{http://arxiv.org/abs/1901.00505}{{\ttfamily
  arXiv:1901.00505 [hep-th]}}.

\bibitem{Chen:2019ror}
H.~Z. Chen and M.~Van~Raamsdonk, ``{Holographic CFT states for localized
  perturbations to AdS black holes},''
  \href{http://dx.doi.org/10.1007/JHEP08(2019)062}{{\em JHEP} {\bfseries 08}
  (2019) 062}, \href{http://arxiv.org/abs/1903.00972}{{\ttfamily
  arXiv:1903.00972 [hep-th]}}.

\bibitem{Martinez:2021uqo}
P.~J. Mart\'\i{}nez and G.~A. Silva, ``{Thermalization of holographic excited
  states},'' \href{http://dx.doi.org/10.1007/JHEP03(2022)003}{{\em JHEP}
  {\bfseries 03} (2022) 003}, \href{http://arxiv.org/abs/2110.07555}{{\ttfamily
  arXiv:2110.07555 [hep-th]}}.

\bibitem{Arias:2020qpg}
R.~Arias, M.~Botta-Cantcheff, P.~J. Martinez, and J.~F. Zarate, ``{Modular
  Hamiltonian for holographic excited states},''
  \href{http://dx.doi.org/10.1103/PhysRevD.102.026021}{{\em Phys. Rev. D}
  {\bfseries 102} no.~2, (2020) 026021},
  \href{http://arxiv.org/abs/2002.04637}{{\ttfamily arXiv:2002.04637
  [hep-th]}}.

\bibitem{Anous:2016kss}
T.~Anous, T.~Hartman, A.~Rovai, and J.~Sonner, ``{Black Hole Collapse in the
  1/c Expansion},'' \href{http://dx.doi.org/10.1007/JHEP07(2016)123}{{\em JHEP}
  {\bfseries 07} (2016) 123}, \href{http://arxiv.org/abs/1603.04856}{{\ttfamily
  arXiv:1603.04856 [hep-th]}}.

\bibitem{Distler:1998gb}
J.~Distler and F.~Zamora, ``{Nonsupersymmetric conformal field theories from
  stable anti-de Sitter spaces},''
  \href{http://dx.doi.org/10.4310/ATMP.1998.v2.n6.a6}{{\em Adv. Theor. Math.
  Phys.} {\bfseries 2} (1999) 1405--1439},
  \href{http://arxiv.org/abs/hep-th/9810206}{{\ttfamily arXiv:hep-th/9810206}}.

\bibitem{Gunaydin:1985cu}
M.~Gunaydin, L.~J. Romans, and N.~P. Warner, ``{Compact and Noncompact Gauged
  Supergravity Theories in Five-Dimensions},''
  \href{http://dx.doi.org/10.1016/0550-3213(86)90237-3}{{\em Nucl. Phys. B}
  {\bfseries 272} (1986) 598--646}.

\bibitem{Khavaev:1998fb}
A.~Khavaev, K.~Pilch, and N.~P. Warner, ``{New vacua of gauged N=8 supergravity
  in five-dimensions},''
  \href{http://dx.doi.org/10.1016/S0370-2693(00)00795-4}{{\em Phys. Lett. B}
  {\bfseries 487} (2000) 14--21},
  \href{http://arxiv.org/abs/hep-th/9812035}{{\ttfamily arXiv:hep-th/9812035}}.

\bibitem{Girardello:1999bd}
L.~Girardello, M.~Petrini, M.~Porrati, and A.~Zaffaroni, ``{The Supergravity
  dual of N=1 superYang-Mills theory},''
  \href{http://dx.doi.org/10.1016/S0550-3213(99)00764-6}{{\em Nucl. Phys. B}
  {\bfseries 569} (2000) 451--469},
  \href{http://arxiv.org/abs/hep-th/9909047}{{\ttfamily arXiv:hep-th/9909047}}.

\bibitem{Skenderis:2002wp}
K.~Skenderis, ``{Lecture notes on holographic renormalization},''
  \href{http://dx.doi.org/10.1088/0264-9381/19/22/306}{{\em Class. Quant.
  Grav.} {\bfseries 19} (2002) 5849--5876},
  \href{http://arxiv.org/abs/hep-th/0209067}{{\ttfamily arXiv:hep-th/0209067}}.

\bibitem{Heller:2011ju}
M.~P. Heller, R.~A. Janik, and P.~Witaszczyk, ``{The characteristics of
  thermalization of boost-invariant plasma from holography},''
  \href{http://dx.doi.org/10.1103/PhysRevLett.108.201602}{{\em Phys. Rev.
  Lett.} {\bfseries 108} (2012) 201602},
  \href{http://arxiv.org/abs/1103.3452}{{\ttfamily arXiv:1103.3452 [hep-th]}}.

\bibitem{Heller:2012je}
M.~P. Heller, R.~A. Janik, and P.~Witaszczyk, ``{A numerical relativity
  approach to the initial value problem in asymptotically Anti-de Sitter
  spacetime for plasma thermalization - an ADM formulation},''
  \href{http://dx.doi.org/10.1103/PhysRevD.85.126002}{{\em Phys. Rev. D}
  {\bfseries 85} (2012) 126002},
  \href{http://arxiv.org/abs/1203.0755}{{\ttfamily arXiv:1203.0755 [hep-th]}}.

\bibitem{Headrick:2014cta}
M.~Headrick, V.~E. Hubeny, A.~Lawrence, and M.~Rangamani, ``{Causality \&
  holographic entanglement entropy},''
  \href{http://dx.doi.org/10.1007/JHEP12(2014)162}{{\em JHEP} {\bfseries 12}
  (2014) 162}, \href{http://arxiv.org/abs/1408.6300}{{\ttfamily arXiv:1408.6300
  [hep-th]}}.

\bibitem{Nakaguchi:2014eiu}
Y.~Nakaguchi, N.~Ogawa, and T.~Ugajin, ``{Holographic Entanglement and Causal
  Shadow in Time-Dependent Janus Black Hole},''
  \href{http://dx.doi.org/10.1007/JHEP07(2015)080}{{\em JHEP} {\bfseries 07}
  (2015) 080}, \href{http://arxiv.org/abs/1412.8600}{{\ttfamily arXiv:1412.8600
  [hep-th]}}.

\bibitem{Skenderis:2009ju}
K.~Skenderis and B.~C. van Rees, ``{Holography and wormholes in 2+1
  dimensions},'' \href{http://dx.doi.org/10.1007/s00220-010-1163-z}{{\em
  Commun. Math. Phys.} {\bfseries 301} (2011) 583--626},
  \href{http://arxiv.org/abs/0912.2090}{{\ttfamily arXiv:0912.2090 [hep-th]}}.

\bibitem{AlBalushi:2020kso}
A.~Al~Balushi, Z.~Wang, and D.~Marolf, ``{Traversability of Multi-Boundary
  Wormholes},'' \href{http://dx.doi.org/10.1007/JHEP04(2021)083}{{\em JHEP}
  {\bfseries 04} (2021) 083}, \href{http://arxiv.org/abs/2012.04635}{{\ttfamily
  arXiv:2012.04635 [hep-th]}}.

\bibitem{Balasubramanian:2023xyd}
V.~Balasubramanian, Y.~Nomura, and T.~Ugajin, ``{de Sitter space is sometimes
  not empty},'' \href{http://arxiv.org/abs/2308.09748}{{\ttfamily
  arXiv:2308.09748 [hep-th]}}.

\bibitem{Balasubramanian:2022gmo}
V.~Balasubramanian, A.~Lawrence, J.~M. Magan, and M.~Sasieta, ``{Microscopic
  origin of the entropy of black holes in general relativity},''
  \href{http://arxiv.org/abs/2212.02447}{{\ttfamily arXiv:2212.02447
  [hep-th]}}.

\bibitem{Balasubramanian:2022lnw}
V.~Balasubramanian, A.~Lawrence, J.~M. Magan, and M.~Sasieta, ``{Microscopic
  origin of the entropy of astrophysical black holes},''
  \href{http://arxiv.org/abs/2212.08623}{{\ttfamily arXiv:2212.08623
  [hep-th]}}.

\bibitem{Bzowski:2022kgf}
A.~Bzowski, ``{Wormholes, geons, and the illusion of the tensor product},''
  \href{http://arxiv.org/abs/2212.10652}{{\ttfamily arXiv:2212.10652
  [hep-th]}}.

\bibitem{Witten:2021nzp}
E.~Witten, ``{A Note On Complex Spacetime Metrics},''
  \href{http://arxiv.org/abs/2111.06514}{{\ttfamily arXiv:2111.06514
  [hep-th]}}.

\bibitem{Headrick:2009pv}
M.~Headrick, S.~Kitchen, and T.~Wiseman, ``{A New approach to static numerical
  relativity, and its application to Kaluza-Klein black holes},''
  \href{http://dx.doi.org/10.1088/0264-9381/27/3/035002}{{\em Class. Quant.
  Grav.} {\bfseries 27} (2010) 035002},
  \href{http://arxiv.org/abs/0905.1822}{{\ttfamily arXiv:0905.1822 [gr-qc]}}.

\bibitem{Hohenberg:1977ym}
P.~C. Hohenberg and B.~I. Halperin, ``{Theory of Dynamic Critical Phenomena},''
  \href{http://dx.doi.org/10.1103/RevModPhys.49.435}{{\em Rev. Mod. Phys.}
  {\bfseries 49} (1977) 435--479}.

\end{thebibliography}\endgroup

\end{document}